\documentclass{kapedbk}

\usepackage{edbkps}
\normallatexbib
\usepackage{epsfig}
\begin{document}
\def\lsim{\mathrel{\raise.4ex\hbox{$<$}\kern-0.8em\lower.7ex\hbox{$\sim$}}}

\articletitle{Charge Inhomogeneities in Strongly \\ Correlated Systems}
\articlesubtitle{}
\author{A.\ H.\ Castro Neto}
\affil{Department of Physics, Boston University\\
Boston, MA, 02215, USA}
\email{neto@bu.edu}
\and 


\author{C.\ Morais Smith}
\affil{D\'epartement de Physique, Universit\'e de Fribourg, P\'erolles\\
CH-1700 Fribourg, Switzerland}
\email{cristiane.demorais@unifr.ch}


\begin{abstract}
We review the problem of stripe states in strongly correlated systems,
and some of the theoretical, numerical and experimental methods used in 
the last few years to understand these states. We compare these states
to more traditional charge-ordered states such as charge density waves (CDW)
and phase-separated systems. We focus on the origin of 
stripe states as an interplay between magnetic and kinetic energy, and
argue that the stripe state is generated via a mechanism of kinetic
energy release that can be described via strongly correlated models such
as the t-J model. We also discuss phenomenological models of stripes,
and their relevance for magnetism and for the pinning of stripes by the
underlying lattice and by disordered impurities.
Recent experimental evidence for the existence of
stripe states in different cuprate systems is also reviewed.
\end{abstract}
\section{Introduction}

The problem of strong electron correlations in 
transition-metal oxides and U and Ce intermetallics 
has been a subject of intense research in 
the last 20 years. This interest has been driven mostly by puzzling
experimental findings in materials such as organic conductors
and superconductors, heavy-fermion alloys, and high-temperature
superconductors. These systems
are characterized by large Coulomb interactions, low dimensionality,
strong lattice coupling,
and competition between different phases: antiferromagnetism,
ferromagnetism, spin density waves (SDW), superconductivity,
and charge density waves (CDW). The strong interplay between
different order parameters is believed to lead to charge and
spin inhomogeneities, and to a myriad of energy and length scales
that makes the problem very difficult to treat with the
methods and techniques used for the study of Fermi-liquid 
metals. Although the problem of the quantum critical
behavior of metals in the proximity of an isolated zero-temperature 
phase transition has been subject of much study and heated debate \cite{sachdev},
our understanding of the problem of electrons close to multiple
phase transitions is still in its infancy.  
Here we will review both the experimental
evidence for the existence of certain inhomogeneous states 
called "stripe" states and 
some of the current theoretical approaches used to understand their origin, 
nature, and importance in the context of magnetism and superconductivity
\cite{erica_review}.

The formation of static or dynamic spin-charge stripes in strongly
correlated electronic systems has been corroborated recently 
by several experiments, especially in manganites \cite{mang,STMmang},
nickelates 
[5-8], and cuprates [9-34].
The experiments span a large variety of techniques, from 
scanning tunneling microscopy (STM) [4,9-14], 
neutron (and x-ray) scattering [5,6,15-25],
nuclear magnetic (and quadrupole)
resonance (NMR and NQR) \cite{nickNMR,NMR2,NMR1}, muon spin rotation 
($\mu$SR) \cite{muSR,ako},
optical and Raman spectroscopy \cite{nickOpt,Opt,basov}, 
transport \cite{noda}, angle-resolved
photoemission (ARPES) \cite{ARPES-LSCO}, and ion channeling \cite{Venky}. 

Charge and spin modulated states, such as CDWs, Wigner crystals, SDWs,
antiferromagnetism, and ferrimagnetism are common occurrences in many 
transition-metal compounds. These systems are characterized by an order parameter (such
as the charge and/or spin density) that is modulated with a well-defined
wave vector ${\bf Q}$. Because of the modulations and the coupling to
the lattice, these states usually present lattice distortions 
which are easily observed in diffraction 
experiments such as neutron scattering. In this regard the stripe states discussed
here are very much like CDW/SDW instabilities, except that in CDW systems
the ordered state is driven by a Fermi-surface instability (usually generated
by nesting and/or Van Hove singularities), and Coulomb effects are 
secondary because of good screening. The clearest example of such Fermi-surface
effects occurs in Cr alloys, where the system undergoes a phase transition into
a CDW/SDW state \cite{cr}. The CDW and SDW transitions occur 
at the {\it same temperature}, and the charge
order has a period that is $1/2$ that of the SDW. The main difference
between the phase transition in Cr and the stripe states to be discussed here is that
the charge order in stripe systems occurs at higher temperature than the spin
order \cite{tran}. 
Thus, on reducing the temperature the onset of charge order occurs first and
the spins simply follow. 
In a weak-coupling analysis of Fermi-surface instabilities, this type of transition
is not possible because
of reconstruction of the Fermi surface due to the appearance of long-range order. 
Thus there are different energy scales for the charge and spin order
that characterizes the materials discussed here. 

We should stress that the CDW and/or SDW instabilities in metallic 
systems are not trivial, and although we understand the basic mechanisms
which drive these instabilities \cite{halperin} 
our knowledge of their origin and effects on the electronic degrees of 
freedom is far from complete.
Systems such as transition-metal dichalcogenides \cite{tmd} 
have a high
temperature CDW transition with a very anomalous metallic phase and show in addition 
the phenomenon of "stripe formation" \cite{stripe_tmd}. 
The stripes in these CDW systems
are understood, however, because the CDW order is incommensurate with the
lattice and therefore phase fluctuations of the CDW order parameter are 
allowed energetically. Local CDW phase-slips give rise to a filamentary
stripe phase which, in fact, has a Fermi-surface origin. 

To understand the origin of CDW stripes one may consider the 
complex order parameter 
$\Delta$ for a CDW with incommensurate ordering wave vector ${\bf Q}$.
The free energy of the problem may be expressed as \cite{mm}
\begin{eqnarray}
F = F_0[\Re(\Delta)] + \int d{\bf r} \left\{
\frac{1}{2 m^* Q^2} \left[
\left|{\bf Q} \cdot (\nabla - i {\bf Q})\Delta\right|^2
+ \kappa |{\bf Q} \times \nabla \Delta|^2 \right]\right\}
\label{fcdw}
\end{eqnarray}
where $F_0[x]$ is a minimal polynomial of $x$ that respects the symmetry of
the lattice and renders the free energy bounded from below. 
For a triangular lattice, for instance, it can be written as \cite{mm} 
\begin{eqnarray}
 F_0[\Re(\Delta)] = a({\bf r},T) \Re(\Delta)^2 + b({\bf r},T) \Re(\Delta)^3
+ c({\bf r},T) \Re(\Delta)^4 ,
\label{poly}
\end{eqnarray} 
where the coefficients of the expansion are smooth functions of 
temperature. In particular for the quadratic term $a({\bf r},T) = a_0({\bf r})
(T-T_{ICDW})$
where $T_{ICDW}$ is the transition temperature of the incommensurate
CDW (ICDW) state. In (\ref{fcdw}) $m^*$ and $\kappa$ are parameters 
specific to the material under consideration 
and the derivative terms are written such that the free energy
of the CDW is minimal when the ordering wave vector lies in the correct
direction and has the correct wave length. In the case of an ICDW 
the order parameter is obtained by minimizing (\ref{fcdw}) to give 
\begin{equation}
\Delta_{ICDW}({\bf r}) = \Delta_{0,I} e^{i {\bf Q} \cdot {\bf r}},
\label{dicdw}
\end{equation}
where $\Delta_{0,I} = \sqrt{2 a_0 (T_{ICDW}-T)/(3 c_0)}$ for $T<T_{ICDW}$
and zero otherwise. The parameter $c({\bf r},T) = c_0 =$ constant.
Here we have assumed that the parameters in
(\ref{poly}) may be expanded in a form such as $b({\bf r}) = b_0 + 
b_1 \exp\{i {\bf K}_i \cdot {\bf r}\}$ where ${\bf K}_i$ are the shortest
reciprocal-lattice vectors characteristic of the lattice symmetry. 

In the commensurate CDW (CCDW) case the wave vector of the order
parameter ``locks'' with the lattice so that its modulation becomes
a fraction of the lattice wave vector ${\bf K}_1$. Then one
would replace ${\bf Q}$ in (\ref{dicdw}) by ${\bf K}_1/q$,
where $q$ is an integer, and $\Delta_0$ by a value $\Delta_{0,C}$ which
must be calculated from the free energy and depends in general
on various coefficients of $F_0$ in (\ref{poly}). The
transition temperature $T_{CCDW}$ is usually smaller than $T_{ICDW}$
so that the generical behavior of the system consists of two transitions, first
into an incommensurate phase and then into a commensurate phase \cite{mm}.
In many systems the ICDW-CCDW transition does not occur
and the system remains incommensurate down to very low temperatures
\cite{tmd}. 

In order to study the problem of the commensurate-incommensurate transition,  
and the topological defects which
appear due to incommensurability, one must generalize (\ref{dicdw})
to include phase fluctuations. These may be incorporated by 
writing the order parameter in the form
\begin{eqnarray}
\Delta({\bf r}) = \Delta_0 e^{i \frac{1}{q} {\bf K}_1 \cdot {\bf r}
+ i \theta ({\bf r})},
\end{eqnarray}
where $\theta({\bf r})$ is the angle variable which 
determines the commensurability
of the system: for the ICDW 
$\theta({\bf r}) = ({\bf Q}-{\bf K}_1/q)\cdot {\bf r}$, while for a CCDW $\theta=0$. Because we are considering
the simplest problem of a single CDW wave vector the problem becomes
effectively one-dimensional if the variables are redefined in a new, 
rotated, rescaled, reference frame defined by 
${\bf s} = (x,y) = |{\bf Q}-{\bf K}_1/q| {\bf r}$. In this case
it is obvious that $\theta({\bf r}) = \theta(x)$, and by direct
substitution of (\ref{fcdw}) we find that the dimensionless free energy 
per unit of length relative to the commensurate case becomes
\begin{eqnarray}
\delta f = \int dx \frac{1}{2} \left\{ \left[ \partial_x \theta(x)-1 \right]^2
  + g  \left[1 - \cos(q \theta)\right]\right\}
\label{sg}
\end{eqnarray}
where $g$ is the coupling constant of the system and depends on the
parameters of (\ref{fcdw}). The free energy in (\ref{sg}) describes
a sine-Gordon model where the cosine term favors
the commensurate state ($\theta(x)=0$) while the gradient term favors
the incommensurate state ($\theta(x) = x$). Thus $\theta$ is the
order parameter and the 
discrete symmetry $\theta \to \theta + 2 \pi/q$ is broken in the
ordered phase. There is therefore 
a critical coupling value $g_c$ that separates these 
two phases. However, it is easy to see that there are other solutions
which minimize the free energy. In fact, variation of 
(\ref{sg}) with respect to $\theta$ yields 
\begin{eqnarray}
\frac{d^2 \theta}{d x^2} = g \, q \,  \sin(q \theta)
\label{sge}
\end{eqnarray}
which has a particular solution
\begin{eqnarray}
\theta_K(x) = \frac{2}{q} \arctan\left(e^{\sqrt{g} x}/2\right),
\label{onekink}
\end{eqnarray}
where the boundary conditions are $\theta(x=-\infty)=
d \theta(x=-\infty)/dx=0$. Notice that (\ref{onekink}) changes
smoothly from $\theta=0$ at $x=-\infty$ to $\theta = 2 \pi/q$ 
when $x \gg 1/\sqrt{g}$. In the context of the sine-Gordon model
this is called a topological soliton or kink, while in the CDW literature
\cite{mm} it is called a discommensuration. In general, the solution
of (\ref{sge}) is given by \cite{davidov}:
\begin{eqnarray}
\theta(x) = \frac{2}{q} \arcsin[\eta {\rm sn}(\sqrt{g} x/k, k)]
\label{full}
\end{eqnarray}
where $\eta = \pm 1$ and sn$(u,k)$ denotes the sine-amplitude, which is
a Jacobian elliptic function of modulus $k$. The sine-amplitude 
is an odd function of its argument $u$ and has 
period $4 K(k)$, where $K(k)$ is the complete elliptic
integral of the first kind. By substituting (\ref{full}) in (\ref{sg})
and minimizing with respect to $k$ one finds that $k \approx 
2 \sqrt{g}$. In the limit $g \to 0$ one has also $k \to 0$ and 
sn$(u,k) = \sin(u)$, whence $\theta(x) \approx x$  
as expected. On the other hand, when $g \gg 1$ one finds $k \approx 1$
and sn$(u,k) \approx \tanh(u)$, in which case (\ref{onekink}) is obtained. 
For a generic value of $g$ the solution has the form of a staircase.
The plateaus in the staircase are multiples of $2 \pi/q$ and correspond
to regions where the CDW is in phase with the lattice, while in the
transitions between the plateaus the CDW is not locked, leading
to discommensurations. For large values of $g$ the discommensurations
are rather narrow and we find stripe states. These states have
been observed experimentally in CDW systems \cite{stripe_tmd}.

In contrast to these CDW stripes, the stripe systems which we will discuss here
have their origin in Mott insulators with very large Coulomb energies, whereas typical
CDW/SDW systems are very good metals in their normal phase (Cr is a shiny
metal while La$_2$CuO$_4$ is opaque and grayish). It is exactly
the ``mottness'' of these systems which complicates the theoretical understanding of 
their nature. If we take seriously the analogy between dichalcogenides 
and cuprates we could think of the stripes as phase-slips of an incommensurate
order parameter associated with the Mott phase. The primary question concerns
the order parameter of a Mott phase.  Antiferromagnetic order 
usually occurs in a Mott insulator but is essentially a
parasitic phase (systems of spinless electrons at 1/2-filling with strong
next-nearest neighbor repulsion, and 
frustrated magnetic systems, can be Mott insulators without 
showing any type of magnetic order \cite{mott_frustrated}). 
Unfortunately, the order
parameter which characterizes the Mott phase in a finite number $d$ of  
spatial dimensions is not known. 
Within the dynamical mean-field theory
($d \to \infty$) the order parameter of the Mott transition
has been identified with a zero mode of an effective Anderson model \cite{gabi}, 
but generalizations for the case of finite dimensions have not been
established. The search for the order parameter which 
characterizes ``mottness'' is one of the important problems of modern condensed
matter physics.

The aim of this work is to summarize the current literature on the stripe 
phase in high-$T_c$ superconductors. Although there is a consensus for the 
existence of stripes in manganites and nickelates, no agreement has yet been 
achieved concerning the superconducting cuprates.
Despite this controversy, the presence of stripes is now firmly
established in La$_{2-x}$Sr$_x$CuO$_4$ (LSCO) 
\cite{tran,yama,Opt,basov,noda,ARPES-LSCO}.
In addition, recent experiments suggest that they may also be present in
YBa$_2$Cu$_3$O$_{7-\delta}$ (YBCO) 
[19-26,34], as well as in
Bi$_2$Sr$_2$CaCu$_2$O$_{8+\delta}$ (BSCCO) \cite{hoff,Kapitulnik}. 
In this work 
we discuss some of the different experimental techniques
which prove or suggest the presence of stripes in cuprates, 
and we present some theoretical ideas on the existence and
relevance of the stripe state. 

This review is structured as follows: in section II, a 
survey of the theoretical derivations of the stripe phase as a ground state 
of models which are appropriate for describing doped Mott insulators is presented. 
In section III we discuss the experimental observation of stripes, and in section 
IV the role of kinetic energy. In section V we consider some 
phenomenological models, which provide a means to go beyond the question of 
existence of stripes and allows one to predict measurable quantities. 
Finally, in section VI, we present the most recent experimental results for
YBCO and BSCO, as well as the open questions and topics of debates. 
In section VII we draw our conclusions. 

\section{Origin of stripes}
 
One of the main problems in condensed matter theory since the
discovery of high temperature superconductors in 1986 \cite{discovery}
is related with the possible dilute phases of Mott insulators \cite{mott}.
These materials have a large charge transfer gap, so at half-filling are 
insulating two-dimensional (2D) antiferromagnets well described by the 
isotropic Heisenberg model \cite{chn}. These are trademarks
of ``mottness'' \cite{philips} and led 
Anderson \cite{anderson} to propose that cuprates may be well 
described by a Hubbard model with large intra-site repulsion
$U$. Later studies showed that close to half-filling and infinite
$U$ the model maps into the $t$-$J$ model \cite{zhang-rice}
\begin{eqnarray}
H = -t \sum_{<i,j>,\alpha} P c^{\dag}_{i,\alpha} c_{j,\alpha} P + 
J \sum_{<i,j>} {\bf S}_i \cdot {\bf S}_j,
\label{tj}
\end{eqnarray}
where $t$ is the hopping energy and $J \approx 4 t^2/U \ll t,U$ is
the exchange interaction between neighboring electron spins, 
${\bf S}_i= c^{\dag}_{i,\alpha} \vec{\sigma}_{\alpha,\beta} c_{i,\beta}$
($c_{i,\alpha}$ is the electron annihilation operator at the site $i$ with
spin projection $\alpha=\uparrow,\downarrow$, and $\sigma^a$ with 
$a=x,y,z$ is a Pauli matrix). In Eq.\ (\ref{tj}) $P$ is the
projection operator onto states with only single site occupancy (double occupancy is
forbidden). Eq.\ (\ref{tj}) reduces trivially to the Heisenberg model at half-filling
and describes the direct interplay between the two main driving
forces in the system, magnetism (characterized by $J$) and kinetic energy 
(characterized by $t$).

\begin{figure}[ht]
  \vskip.2in
  \centerline{\epsfxsize=0.80\columnwidth\epsffile{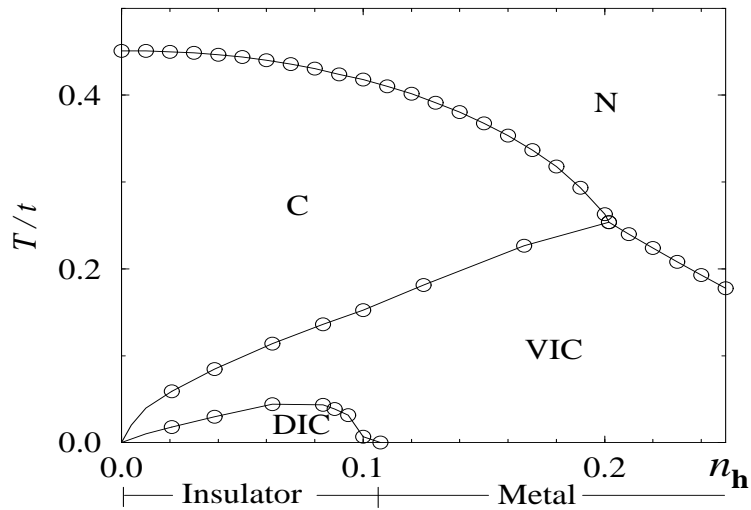}}
  \caption{Phase diagram in the plane of temperature $T$ and hole concentration $x$ 
($n_h = 2x$ for LSCO) obtained
by Machida and Ichioka \cite{machida} from mean-field studies of the Hubbard model.
In the figure, N indicates the normal phase and C denotes the commensurate
antiferromagnetic phase. In addition, two other incommensurate phases exist, with
vertical (VIC) or diagonal (DIC) stripe order. The VIC phase is metallic, but the
DIC is insulating at zero temperature. }
\label{fig1}
  \end{figure}

The existence of a stripe phase in cuprates was first suggested 
in the context of Hartree-Fock studies of the 
Hubbard model close to half-filling and at zero temperature $(T=0)$ \cite{zaanen}. 
This calculation is essentially
analogous to the one used to study CDW/SDW transitions in metallic
materials. For $U < t $, vertical stripes (parallel to the $x$- or $y$-axis
of the crystal)
were shown to be lower in energy \cite{zaanen,schulz,innui}, 
whereas for large $U>t$ diagonal stripes were found to be 
energetically more favorable \cite{innui,poilb}.
The crossover from vertical to diagonal stripes was 
calculated numerically to occur at $U / t \sim 3.6$ \cite{innui}. 
Such calculations have recently been generalized to finite temperatures, and
the phase diagram was derived as a function of $T$ and doping $n_h$, (Fig.\ \ref{fig1} 
\cite{machida}). One important feature of these mean-field calculations is
that they predict the formation of charge-ordered domain walls at which the staggered
magnetization changes phase by $\pi$.
The magnetic order parameter is therefore maximal not at $(\pi/a,\pi/a)$ as in
an ordinary antiferromagnet, but at an incommensurate vector $(\pi/a\pm \delta,\pi/a)$
where $\delta = \pi/\ell \ll 1$ and $\ell$ denotes the charge stripe spacing. This 
incommensurability is an important feature
of the stripe problem because, as we will demonstrate below, it leads to
a reduction in the {\it kinetic} energy of the holes. These calculations,
however, always predict that the stripe states possess a gap. The simple reason
for this effect is that the only way in Hartree-Fock to reduce the energy
of the system is by opening a gap at the Fermi surface. Furthermore, Hartree-Fock
calculations in strongly interacting systems are not quantitatively reliable
because they are unable to take fluctuation effects into account, and therefore
should be considered only as providing some qualitative insight into the
ground-state properties. 
They, however, do provide a "high energy" guide (snapshot picture) 
of the possible phases of the problem, and in fact they have been fundamental
for the interpretation of certain experiments such as neutron scattering. 

Recent theoretical efforts have focused not only on determining
the ground-state properties of the Hubbard, but also of the $t$-$J$ model. Numerical 
techniques such as density matrix renormalization group (DMRG) \cite{DMRG}, quantum 
Monte Carlo \cite{QMC}, and exact diagonalization \cite{ED}, have been applied
to the 2D $t$-$J$ model with differing degrees of success. The main problem 
is that a strongly interacting problem like the $t$-$J$
model is subject to strong finite-size and boundary-condition effects which 
are difficult to control. 

\subsection{Numerical Studies}

Early numerical calculations on the $t$-$J$ model have shown that  
for physical values of $J/t$ and close to half-filling there is a tendency for
phase separation \cite{lek}. This phase
separation can be pictured in the limit of $t=0$ (classical limit)
as a lowering of the system energy by placing all the holes
together in order to minimize the number of broken antiferromagnetic bonds.
This simple picture leads to separation into two distinct phases: a
commensurate, insulating region and an insulating, hole-rich region. 
It naturally overestimates the importance of the magnetic energy
relative to the kinetic energy, and therefore can be correct only when $J \gg t$.
For finite values of $t$, the hole wave function delocalizes and
this picture breaks down. The main question is for which values of $J/t$ a 
phase separation may arise. Emery, Kivelson and Lin \cite{lek} found that 
phase separation can occur for infinitesimal values of $J/t$ sufficiently close to 
half-filling. These results, however, have been questioned
in the light of more recent numerical data. There is no doubt 
that the $t$-$J$ model undergoes phase separation for $J \gg t$ as all 
numerical calculations indicate. Close to the physical region of $J<t$, 
the current evidence for phase separation is weak, and so the issue 
remains controversial. 

DMRG calculations in large clusters \cite{DMRG} indicate the presence of stripe 
correlations
in the $t$-$J$ model. These studies, however, have been criticized on the basis of
the special role of boundary conditions. Recent 
work on the Ising $t$-$J$ model indicate that stripe formation does occur in this
system, independent of the boundary conditions \cite{sasha_stripe}. 
It was shown
via non-perturbative analytical calculations that minimization of the hole
kinetic energy is the driving force behind stripe formation. This
result has been confirmed by a number of numerical calculations 
in the $t$-$J$ model \cite{george}, as well as in the $t$-$J_z$ model 
\cite{uswhite}.
Another important
conclusion from these studies is that in the stripe phase the 
superconducting correlations are extremely weak. In fact DMRG calculations show 
that the stripe state
is a CDW/Luttinger-liquid state with vanishing density of states at the chemical
potential, and thus is naturally insulating \cite{DMRG}. 
The DMRG calculations then 
support the idea that stripe correlations compete with superconductivity instead
of enhancing it. This is consistent with experimental findings in
Nd-doped LSCO, where the superconducting transition is reduced when static stripe
order sets in \cite{tran}.

Although the DMRG calculations were originally performed in an 
$t$-$J$ model, essentially the same physics is found in the $t$-$J_z$ model.
The reason for the similarity between the $t$-$J$ (where the spins are 
dynamical) and the $t$-$J_z$ (where the spins are static) may be understood 
on the basis of the fluctuation timescales for
each component in the problem. In the $t$-$J$ model the 
spins fluctuate with a rate
$\tau_s \approx \hbar/J$, while the timescale for hole motion is 
$\tau_h \approx \hbar/t$. When $J/t <1$ (the physical regime of the model) 
one has $\tau_s > \tau_h$,
that is, the holes move ``faster'' than spins. In this case a
Born-Oppenheimer approximation is reasonable, since the spins
have slow dynamics, and the two problems become essentially identical 
\cite{ED}. 
The advantage of working with the $t$-$J_z$ model is that many
of its properties are significantly easier to study both numerically and 
analytically.

The introduction of next-nearest-neighbor hopping $t'$
favors mobile $d$-wave pairs of holes for $t'>0$, 
and single-hole excitations (spin-polarons) for $t'<0$ \cite{tprime}.
At $t'=0$ the stripe state is very close in energy to the $d$-wave
pair state, and thus a small change in the boundary conditions or inclusion
of small perturbations in the Hamiltonian can easily favor one
many-body state relative the other. The quasi-degeneracy of different
many-body states is an important characteristic of strongly correlated
systems. 
Moreover, in real materials other effects may also be responsible for the
selection of the ground state, that is, for lifting of the
quasi-degeneracy. Indeed, by including  
lattice anisotropies, which arise in the low-temperature
tetragonal (LTT) phase of
LSCO co-doped with Nd, the stripe state can be easily selected 
\cite{branko,normand,becca}:
Hartree-Fock calculations of the Hubbard model have shown that
a very small anisotropy (on the order of ten percent) in the hopping parameter $t$ 
is already sufficient to stabilize the striped phase, independent of the 
boundary conditions (open or periodic) \cite{normand}. 
Monte Carlo studies of the $t$-$J$ model have also confirmed these results 
\cite{becca}.

\subsection{Stripes and phase separation}

\begin{figure}[ht] 
\centerline{\epsfxsize=0.60\columnwidth\epsffile{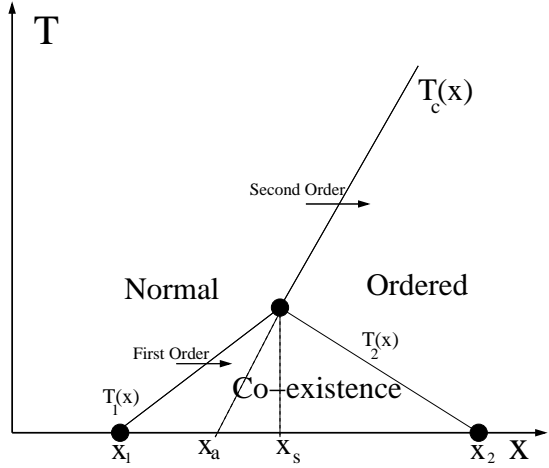}}
\caption{Temperature-doping phase diagram 
for a system close to phase separation. The symbols are
explained in the text.}
\label{pha-sep} 
\end{figure}

The problem of the formation of inhomogeneous states in a system
with phase separation can be easily understood from a classical
point of view by studying the Ginzburg-Landau free energy functional. 
Let $\psi$ be the order parameter of a system described by a
free energy, $F$, of the form 
\begin{eqnarray}
F[\psi] = \alpha(T,x) |\psi|^2 + \frac{\beta(x)}{2} |\psi|^4 
+ \frac{\gamma}{3} |\psi|^6
\label{fps}
\end{eqnarray}
where $\psi$ may be complex for a superconductor, $\alpha(T,x)$, 
$\beta(x)$, and $\gamma>0$ are functions of the temperature
$T$ and some parameter $x$ (such as doping or pressure). 
In the theory of second-order phase transitions, 
the $|\psi|^6$ term is neglected close to the critical line
because $\beta>0$ in this region of the parameter space. Here, however, we
assume that $\beta$ may be negative, and therefore
this term is required so that the free energy is bounded from below.  
The critical line in the $(T,x)$ plane is given by $T_c(x)$
(as shown in Fig.\ \ref{pha-sep}) and we assume the existence
of  a quantum critical point (QCP) at $x=x_a$ (that is, $T_c(x_a)=0$). 
Close to the critical line we introduce the parameterization:
\begin{eqnarray}
\alpha(T,x) &=& \alpha_0 \left[T/T_c(x)-1\right],
\nonumber
\\
\beta(x) &=& \beta_0 \left[T_c(x)/T_c(x_s)-1\right],
\label{glpar}
\end{eqnarray}
while $\gamma$ is approximately independent of $x$ and $T$. 
Notice that with this choice the parameter $\beta(x)$ is positive
for $x>x_s$, signaling that in this regime the transition is
of second order. However, $\beta(x)$ vanishes at $x=x_s$ and
becomes negative for $x<x_s$, indicating that the nature of the
phase transition changes at small $x$. In fact the point
$(x_s,T_c(x_s))$ is a {\it tricritical} point.

For $x>x_s$ the $|\psi|^6$ term is irrelevant close to the phase
transition, and the transition is of second order depending on wether 
$T$ is greater or smaller than $T_c(x)$. Minimizing the free
energy with respect to the order parameter yields 
\begin{eqnarray}
|\psi_0(x,T)|^2 \approx \frac{\alpha_0 \left(1 - T/T_c(x) \right)}{
\beta_0 \left(T_c(x)/T_c(x_s)-1\right)}
\end{eqnarray}
for $T<T_c(x)$ and $x>x_s$. Notice, however, that at $x \lsim x_s$ 
the parameter $\beta(x)$ vanishes, and one must include the $|\psi|^6$ term.
In this case, the free energy has two minima (instead of one) at the critical line 
indicating that the system has two phases, one with $\psi=0$ (normal)
and another with $\psi=\psi_0$ (ordered). 
Minimization of $F$ with respect to the order parameter provides the condition for the 
phase transition 
\begin{eqnarray}
\beta^2(x) &=& \frac{16}{3} \alpha(T^*,x) \gamma \, ,
\nonumber
\\
T^*(x) &=& T_c(x) + \frac{3 \beta^2_0 T_c (x)}{16 \alpha_0 \gamma} 
\left[ \frac{T_c(x)}{T_c(x_s)}-1 \right]^2. 
\label{bag}
\end{eqnarray}
The solution of these equations gives two critical lines,
$T_1(x)$ and $T_2(x)$ on Fig.\ref{pha-sep}. These lines terminate at 
$x_1$ and $x_2$, and for $x_1 < x <x_2$ there is a coexistence region
with two phases (normal and ordered). 

Long-range interactions are readily introduced by modifying the
$|\psi|^4$ term in the free energy to
\begin{eqnarray} 
F_C &=& \int d{\bf r} \int d{\bf r'} |\psi({\bf r})|^2 \frac{e^2}{|{\bf r}-
{\bf r'}|}|\psi({\bf r'})|^2 .
\end{eqnarray}
In this case of fully phase-separated states the cost in electrostatic energy 
is too high and
phase separation is frustrated to a finite length scale, $\ell_{\rm PS}$, that 
depends on the coefficient of the $|\psi|^6$ term. The formation of finite droplets 
with $\psi=\psi_0$ and size $\ell_{\rm PS}$ is therefore more favorable
than the separation of the system into two homogeneous phases with $\psi=0$ and 
$\psi=\psi_0$. Stripes can also be generated in this model if 
one adds terms which break the rotational symmetry 
\begin{eqnarray}
F_{\pm} = \sum_{{\bf q}} \left[ \cos(q_x) \pm \cos(q_y)\right]  |\psi({\bf q})|^2 \, ,
\label{sd}
\end{eqnarray}
depending on whether the interaction with the lattice may be represented in 
terms of $L=0$ (plus sign) or $L=2$ (minus sign) 
angular momentum states ($s$- and $d$-wave, respectively). In the
$L=0$ case a checkerboard state is favored, but even a small $d$-wave term
generates stripes along the $x$- or $y$-directions. 

Disorder can also frustrate the phase separation as one may show 
by adding a ``random mass'' term 
\begin{eqnarray}
F_D = \int d{\bf r} \, \, m({\bf r}) |\psi({\bf r})|^2
\end{eqnarray}
to the free energy, where $m({\bf r})$ is a gaussian variable with average zero
and variance $u$. Using the replica-technique with 
$n$ replicas ($n \to 0$ at the end of the calculation)
and averaging over disorder gives 
\begin{eqnarray}
F = \sum_{a=1}^n F^{(0)}_a - u \sum_{a,b=1}^n \int d{\bf r}  |\psi_a({\bf r})|^2
|\psi_b({\bf r})|^2, 
\end{eqnarray}
where $F^{(0)}_a$ is the free energy without disorder and with $n$ fields $\psi_n$.
Notice that in the replica-symmetric case ($\psi_n=\psi$ for all $n$) the
disorder generates a term of the order $-u |\psi|^4$ which decreases the
effective value of $\beta$, therefore reducing the value of $T_c(x_s)$.
In a renormalization-group (RG) sense this term is relevant, and if the
disorder is sufficiently strong it will bring the tricritical point to
zero temperature (that is, the tricritical point becomes a quantum critical point
(QCP) and will
completely destroy the first-order phase transition). However, in
the ordered phase the system may still possess a coexistence phase with
different values of the order parameter. Once again we would have a
situation where the system forms droplets of the paramagnetic phase
($\psi=0$) inside the ordered phase. The size of these droplets 
depends on the strength of the disorder and it is easy to make them adopt
a stripe conformation by adding a term of the form (\ref{sd}) that
breaks the rotational symmetry.

Although the phenomenology of the problem is quite clear, what is not so evident
is how to apply this theory to the cuprates. Emery {\it et al.} \cite{beg}
proposed that a model similar to the one discussed here (the Blume-Emery-Griffiths
model) may be applied to the cuprates if one defines a pseudo-spin $S_i$ which takes
the values $S_i = +1$ and $S_i=-1$ on regions corresponding to hole-rich and
hole-poor, respectively, whereas $S_i=0$ indicates a local density equal to the 
average value. 
In this case $\psi({\bf r})$ is the coarse-grained version of $S_i$ and
the above discussion is applicable. Notice, once again, that this model completely
disregards the kinetic energy of the problem and can be applied only in the 
situation where $t=0$. It is therefore not at all surprising that 
stripes appear. The inclusion of itinerant degrees of freedom is not straightforward.
One of the main effects of the presence
of itinerant degrees of freedom is the generation of dissipation which can
change the dynamical properties (and exponents) of the system. This problem
has been the object of recent intensive study in the context of quantum 
phase transitions when the coupling between the magnetic order parameter and
the electrons is weak, and the electrons may be treated as a Fermi liquid \cite{hertz}.
In this case the electronic system serves as a heat bath for the relaxation
of the magnetic order parameter.  
In Mott insulators, the mere existence of a Fermi surface and a Fermi-liquid
state can be questioned and thus it is clear that the weak-coupling
formalism cannot be applied to these systems. The charge degrees of freedom
cannot be modeled purely as a heat bath because their feedback effect in the
magnetic system is very strong.

It is interesting to compare the two mechanisms for stripe formation
quoted previously. In one mechanism (represented by the Hartree-Fock calculations), 
stripes are long period 
CDWs arising from Fermi-surface nesting in a weakly incommensurate system 
\cite{zaanen,schulz,innui,poilb}.
There are four features which arise
from a Fermi-surface instability: 1) the transition is spin driven, {\it i.e.,}
there is a single transition temperature $T_c$ 
below which the broken-symmetry solution
of the Hartree-Fock equations is stable; 
2) in the low-$T$ phase there
are gaps or pseudo-gaps on the Fermi surface; 3) the spacing between domain
walls is equal to $\pi/x$, where $x$ is the hole concentration; 4) 
the high-$T$ phase should be a Fermi liquid.

In the case of stripes arising from Coulomb-frustrated phase separation, the
situation is quite different \cite{emery}:
1) the transition is charge driven, {\it i.e.,}
local spin order between the antiphase domain walls can develop only after
the holes are expelled from the magnetic regions. Ginzburg-Landau considerations
indicate either a first-order transition, in which spin and charge order arise 
simultaneously, or a sequence of transitions in which first the charge
order and then the spin order appears as $T$ is lowered \cite{pryadko}; 
2) the stripe spacing
is not necessarily a simple function of $x$, and there is no reason to expect
the Fermi energy to lie in a gap or pseudo-gap; 3) a high-$T$ Fermi-liquid
phase is not a prerequisite.

\section{Experimental detection of stripes in high-$T_c$ cuprates: 
LSCO}

Although the first predictions for stripe formation in doped Mott insulators
were made 13 years ago, not much attention was paid to these results
in connection with high-$T_c$ superconductors 
until 1995, when experimental data from neutron-scattering measurements
in cuprates were interpreted consistently within a stripe picture \cite{tran}.
Co-doping of cuprates has been extremely important for  
revealing the modulated charge states. However, the inclusion of
co-dopant usually reduces $T_c$, raising doubts about the coexistence
of superconductivity and the striped phase \cite{tran,buechner}.  
The first experimental detection of stripes in the cuprates was achieved
in a Nd co-doped compound La$_{2-x-y}$Nd$_y$Sr$_x$CuO$_4$. For $y=0.4$ 
and $x=0.12$, Tranquada {\it et al}.\ \cite{tran} found that the
commensurate magnetic peak at ${\bf Q} = (\pi/a, \pi/a)$ shifts by a 
quantity $\delta= \pi/\ell $, giving rise to four incommensurate peaks. In addition,
new Bragg peaks appear at the points $(\pm 2 \delta, 0)$ and $(0, \pm 2\delta)$, 
indicating that the charges form domain walls 
separated by a distance $\ell $, and that the staggered magnetization undergoes a 
phase-shift of $\pi$ when crossing them. The position of the peaks indicates that the 
stripes are oriented along the
vertical and horizontal directions, with a density of one hole per two Cu
sites (quarter-filled).

The reason why static stripes could be detected in this compound is based 
on a structural transition induced by the Nd atoms. Indeed, co-doping with many
rare-earth species, including Nd and Eu, produces a buckling of the oxygen octahedra 
around the Cu sites and a corresponding transition from the low-$T$
orthorhombic (LTO) to a low-$T$ tetragonal (LTT) phase. The critical concentration
of Nd needed to destroy superconductivity is a function of the charge-carrier
density, {\it i.e.,} the concentration $x$ of Sr atoms. However, the buckling
angle of the octahedra is a universal parameter: for tilts above a critical angle
$\theta \sim 3.6^{\circ}$, superconductivity is completely suppressed in 
these materials \cite{buechner}. For the values of Nd co-doping $y$ used in the
first neutron-scattering experiments, superconductivity was still present, and
the authors claimed that in their samples $T_c \sim 5$ K. However, coexistence 
of static stripe order and superconductivity in LSCO is an issue
that remains controversial, although recent experiments in La$_2$CuO$_{4 + y}$
have shown the coexistence of these two phases \cite{lco} in the same volume
of the sample.

Another important factor assisting the detection of charge stripes in
LSCO systems by elastic neutron scattering was the selected doping
concentration $x \approx 1/8$. The $1/8$ anomaly was known since 1988, when
electrical resistivity measurements in La$_{2-x}$Ba$_x$CuO$_4$ (LBCO) were
first performed \cite{LBCO}. A mysterious reduction of $T_c$ was  
detected around $x = 1/8$, but the understanding of this phenomenon was 
possible only recently, in the light of the stripe picture. Indeed, Ba substitution
also induces a structural transition, similar to rare-earth co-doping, which
probably acts to pin the stripe structure, stabilizing the charge ordering, and hence
reducing $T_c$. Recently, Koike {\it et al.} have shown
that the $1/8$ phenomenon is common to all the cuprates \cite{koike} and 
that a similar effect must occur for $x \sim 1/4$ \cite{1/4}.

\begin{figure}[ht]
\vskip-.8in
  \centerline{\epsfxsize=1.00\columnwidth\epsffile{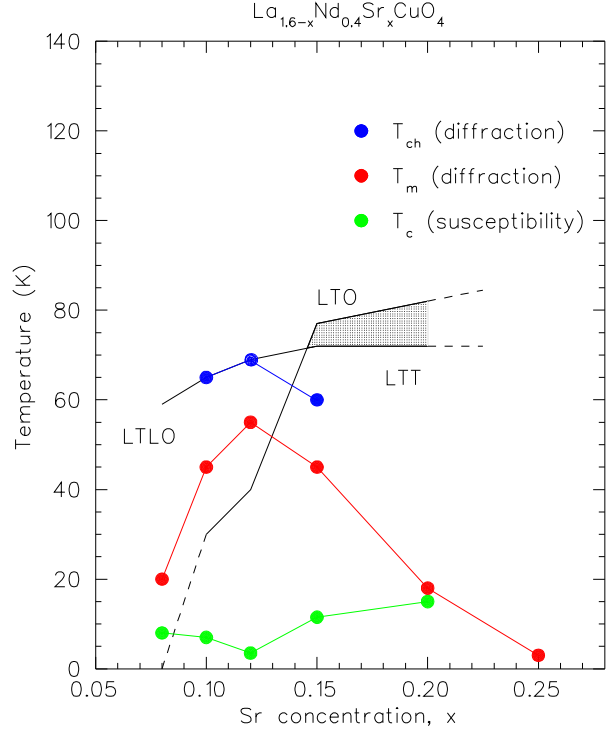}}
 \vskip-1.2in  
\caption{Experimental phase diagram obtained by Ichikawa et al. \cite{tran}
for Nd-doped LSCO. $T_{ch}$  and $T_m$ denote, respectively, the temperatures below 
which charge and spin ordering could be detected in this system by elastic neutron
scattering measurements. The superconducting transition temperature $T_c$ obtained
by susceptibility measurements is also shown. In addition, the structural transition
lines from the low-temperature orthorhombic (LTO) to the low-temperature tetragonal
(LTT) and to the low-temperature-less-orthorhombic (LTLO) phases are displayed.  }
\label{Ichi}
\end{figure}

\begin{figure}[ht]
\vskip-.2in
  \centerline{\epsfxsize=1.00\columnwidth\epsffile{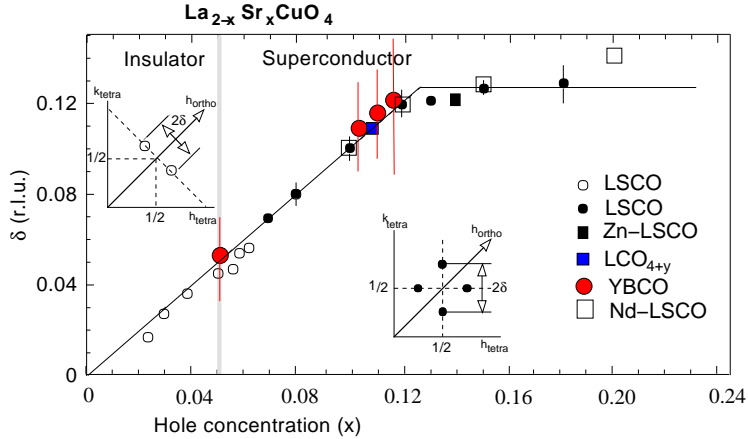}}
\vskip-3.5in  
\caption{Summary of data concerning incommensurability $\delta = \pi/\ell$ as a
function of doping concentration $x$. Data were obtained from neutron-scattering 
measurements by
several groups: open and full small circles are from \cite{SG} and
\cite{yama}, respectively; dark squares are from \cite{hiro}, the grey one
from \cite{yama}, and the white one from \cite{tran}; large circles are from
Refs. [17-22]. }
\vskip-.2in  
\label{IC}
\end{figure}

Though the first neutron scattering experiment was performed in a Nd doped
sample with the ``magic'' hole concentration $x = 1/8$, further measurements
on samples with $x = 0.10$ and $x = 0.15$ confirmed the existence of
incommensurate peaks in the spin and charge sectors, giving support to the
stripe picture (Fig.\ \ref{Ichi} \cite{tran}). Moreover, systematic studies 
of superconducting LSCO samples with a range of doping 
values $x$ has been performed by inelastic neutron scattering \cite{yama}. 
The detected incommensurability is exactly the same as that 
obtained in co-doped samples (Fig.\ \ref{IC} \cite{tran}). 
Both elastic and inelastic neutron scattering, in addition to 
NMR \cite{NMR1}, NQR, $\mu$SR \cite{muSR}, Hall transport \cite{noda}, 
and ARPES \cite{ARPES-LSCO} measurements 
indicate that stripes are present in LSCO. A linear dependence of the 
incommensurability $\delta$ as a function of the doping concentration $x$ has been 
detected for $x < 1/8$, indicating that the stripes behave as ``incompressible" 
quantum fluids in this regime, that is, for $0.05 < x < 0.12$ the 
hole density in each stripe is fixed (one hole per two Cu sites), and by
increasing the amount of charge in the system
one consequently increases the number of stripes and reduces
their average separation $\ell(x)$. Moreover, Yamada {\it et al}.\ 
\cite{Tclinear}
showed that in this regime $T_c$ is also proportional to
$\delta$, {\it i.e.,} $T_c \propto x \propto \delta \propto 1/\ell(x)$. 
Above $x = 1/8$, however, the behavior of the system changes and $\delta$
nearly saturates, indicating a transition to a more homogeneous phase.
Recently, neutron scattering experiments were performed within the
spin-glass regime, for $0.02 < x < 0.05$ \cite{SG}. The result was
surprising: the incommensurate peaks are rotated by 45$^\circ$ in  
reciprocal space, suggesting that the stripes are diagonal and half-filled,
with one hole per Cu site, analogous to nickelate stripes \cite{SG}. 
However, this conclusion may be premature. Because the incommensurate peaks are
observed only in the spin, but not in the charge sector, other explanations
of the phenomenon are plausible, such as the formation of a spiral phase
\cite{nils_glass}. We will discuss this topic below in Secs.\ 4.2 and 5. 
A summary of available experimental
data concerning the incommensurability is presented in Fig.\ \ref{IC}.

It is interesting to compare the two mechanisms proposed theoretically
[49-53,67] for stripe formation in the light of the experimental results \cite{tran}.
Charge order indeed appears before 
spin order, favoring the Emery-Kivelson proposal of frustrated phase separation 
\cite{emery}, but the stripe separation
clearly displays a linear dependence on the inverse of the hole density, 
as predicted by the Hartree-Fock analysis [49-53].
Concerning the stripe filling, the Hartree-Fock predictions are observed
in the spin-glass regime, whereas the Emery-Kivelson proposition holds
within the superconducting underdoped regime.
However, recent slave-boson studies of the 3-band Hubbard model have shown that
if the oxygen-oxygen hopping integral $t_{pp}$ is finite, quarter-filled stripes are
more stable than half-filled ones \cite{loren}. DMRG studies of the $t$-$J$ model 
found also that quarter-filled stripes are the lowest-energy configuration
\cite{DMRG}. 

\section{The single-hole problem: the role of kinetic energy}

The problem of a single hole in a 2D antiferromagnet has a long
history and it is probably one of the best-studied cases of strongly
correlated systems. This problem is by nature single particle
because it deals with a single particle interacting
with a complex magnetic environment. Physically this is  
realized only in physical systems with extremely small carrier densities, 
and many of their properties can be related to the polaron problem \cite{sasha_review}. 
Here we will not review the problem in any way (there are very good
reviews on the subject) but we would like to stress the important role that 
the hole kinetic energy plays in determining the possible
phases. Moreover, as we are going to show,
the same physics may be responsible for the stripe phases in 
transition-metal oxides.

The simplest limit of the $t$-$J$ model with a single hole is the limit
of $J=0$, that is, $U \to \infty$. This limit was studied by Nagaoka
\cite{Nagaoka}, who showed that a single hole makes the system 
unstable toward a ferromagnetic
phase. The origin of ferromagnetism in this case lies in the minimization
of the hole kinetic energy: because double occupancy is precluded, 
the kinetic energy conserves the spin projection, and there is no
energy penalty for the formation of ferromagnetic bonds,
the kinetic energy is minimal when all the spins have the same
direction. This
process is essentially the same as that occuring in double-exchange
systems such as manganites, where a ferromagnetic coupling between 
the electron spin and a magnetic host produces a ferromagnetic state 
by minimizing the kinetic energy. For finite but small $J$
($J/t \ll 1$) the same effect occurs, but instead of polarizing
the entire plane of spins a single hole produces a ferromagnetic
polarization cloud of size $R$: the hole gains a kinetic energy
of order $4 t - t/(R/a)^2$ by being free to move in the ferromagnetic
region but has to pay a magnetic energy cost of order $J (R/a)^{d-1}$
for the generation of a frustrated magnetic surface with ferromagnetic
bonds. Minimization of the total energy of the hole indicates that the
radius of the ferromagnetic region decreases according to $R/a \sim 
(t/J)^{1/(d+1)}/(d-1)$ as $J/t$ increases (notice that for the case $d=1$
this estimate always produces $R = \infty$ for all values of 
$t/J$) \cite{affleck_white}. 

As $J/t$ increases the magnetic
energy generated by the frustrated magnetic surface becomes too large, and
$R$ shrinks to zero. Thus, larger values of $J/t$ lead to a change in the physics.
A simple way to reduce the magnetic frustration is to reduce
the frustrated surface of the spin configuration. Instead of a frustrated
surface of misaligned spins it becomes energetically favorable to create
strings of ferromagnetic pairs of spins due to the retraceable motion of the 
holes. It is clear that in this case the energy of the string grows
linearly with its size $L$, and therefore that the energy required to generate
a string is approximately $J L$. On the other hand, the hole kinetic
energy changes from $-t$ to a quantity of order $-t + t/(L/a)^2$, and the
problem is essentially equivalent to that of a single particle
in a linearly confining potential. The solution of this quantum mechanical
problem is straightforward and minimization of the total energy shows that
the size of the strings varies according to $L \sim (t/J)^{1/3}$ independent
of the dimensionality. Thus, on increasing $J/t$ the
single hole case exhibits a crossover from a ferromagnetic polaron to
the so-called spin-polaron. 
The confinement described here is not completely correct because 
the hole has been considered as a semi-classical entity whereas in fact it 
is a fully quantum-mechanical object, which could 
undergo quantum tunneling over classically
forbidden regions. This tunneling gives rise to ``Trugman loops'' where the 
hole can move diagonally but with a very small tunnel
splitting (that is, very large effective mass) \cite{trugman}. 

In any case, the true situation lies between these extremes and ferromagnetic 
polarization is concomitant with string processes. It is clear
that the problem of the doped antiferromagnet centers on the 
compensation of magnetic frustration by reduction of the kinetic
energy. Furthermore, the string mechanism provides a clear
way to release kinetic energy, namely the retraceable motion of the hole. 
By incorporating both the kinetic energy (creation of ferromagnetic
bonds) and the magnetic energy (generation of strings) one 
may understand how holes can move in a system with strong antiferromagnetic
correlations. We have, however, discussed only the case of a single hole 
but for superconductivity it is important
to understand the situation when the density of holes increases. The first
crossover in these systems occurs when a finite
{\it linear} density of holes (say $N/L$ is finite but $N/L^2$ zero) is reached.
The second crossover occurs when $N/L^2$ becomes finite.
Thus, in such strongly correlated systems one has always at least 2 crossovers:
from single particle to 1D and from 1D to 2D. In the next section we will 
discuss the first crossover and show that it is related to
the formation of stripes.

\subsection{Crossover from single particle to 1D: stripes in the t-J model}

\begin{figure}[ht] 
\centerline{\epsfxsize=0.40\columnwidth\epsffile{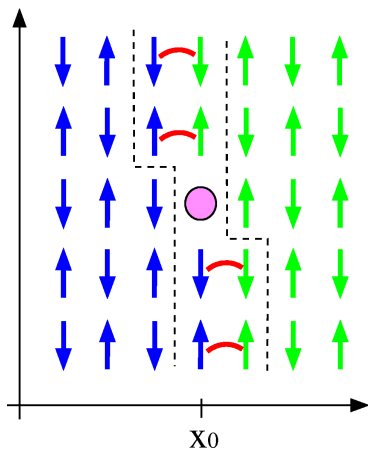}}

\caption{Antiphase domain wall with one hole. Thick lines represent
broken bonds, while dashed lines mark the position of the topological defect.}
\label{apdw}
\end{figure}

Consider an infinite antiphase domain wall oriented along one of the crystal axes
directions of the system (Fig.\ref{apdw}). The cost in energy
per hole to create such a state is $J/2(n^{-1}-1)$, where $n$ is the
linear density of holes along the stripe. However, the hole wave function is
translationally invariant along the stripe and the 
kinetic energy gain due to longitudinal hopping is $-2 t \sin(\pi n)/(\pi n)$.
For $J/t=0.4$ one may show that the energy is minimized
for $n=0.32$ with an energy $E_b \approx -1.255 t$ which is larger than
the energy of a hole in the bulk (spin polaron), given by $E_{sp} = -2 \sqrt{3} t
\approx -2.37 t$ \cite{ED}. 
Here we have not included 
the transverse motion of the hole perpendicular to the stripe, which 
further reduces the kinetic energy of the system but 
gives also a finite width to the hole wave function. Using a retraceable-path
approximation (but ignoring hole-hole interactions) 
one may calculate analytically the Green function for
the holes \cite{sasha_stripe}. 

Holes are confined to an
antiphase domain wall by the potential generated by
strings of overturned spins 
(that is, there is a linearly growing potential
transverse to the stripe direction). One may show also that in
this configuration the hole is actually a holon, {\it i.e.,} it
carries charge but no spin and any motion of the hole away
from the stripe produces a spinon, a particle with a spin of 1/2 but no charge, 
of energy $J$. In the bulk the
hole carries both spin and charge and therefore is a spin-polaron.
Spin-charge separation is thus local, but not macroscopic 
\cite{sasha_stripe}. Because
of this effect, Trugman loops \cite{trugman}, which are
responsible for hole deconfinement in the absence of antiphase
domain walls, are not effective because the motion
of holes away from the wall always produces an excitation of
finite energy. 

One finds that for $J/t=0.4$ and $n \approx 0.3$ the
energy of the stripe state is $E_0 \approx -2.5 t$, and therefore lower
than the energy of the spin polaron. Furthermore, the
width of the hole wave function has a value on the order of 3 - 4 lattice units
\cite{sasha_stripe}, and therefore extends a considerable distance from the
antiphase domain wall, in contrast to the ``cartoon'' picture
where the stripe has a width of only one lattice spacing \cite{tran}. 
Thus, considering a stripe as a completely 1D
object is somewhat misleading because each hole may make long incursions into
the antiferromagnetic regions. Moreover, it is clear from these analytic calculations
that it is the single-hole kinetic energy which is
responsible for the stabilization
of the stripe state. These results have been confirmed numerically by DMRG
and exact-diagonalization studies \cite{george}. In previous works semi-phenomelogical
field theoretical  models were proposed to explain the formation of anti- and 
in-phase domain walls and stripes \cite{antonioz,zachar,priadko}. However, from the
studies on the $t$-$J_z$ model it becomes clear that stripe formation is a 
short distance problem (that is, it involve high energy states) and cannot
not be properly addressed with the use of field theories that can only 
describe the low-energy, long-wavelength physics. 

We should stress that we are discussing the 
ground state, that is, the lowest-energy 
stationary state and therefore
the concept of ``stripe fluctuations" refers, in this 
context, to excitations which are separated from the ground
state by an energy of order $t(J/t)^{2/3}$ because of
confinement in the transverse direction \cite{sasha_stripe}. 
These results are consistent not only with the 
DMRG results for the $t$-$J$ model for small $J/t$ \cite{george}
but also with those for the $t$-$J_z$ model 
\cite{uswhite}. As a consequence the stripe is metallic, in contrast 
to Hartree-Fock results which always produce a gap \cite{hf}. 
As in a Luttinger liquid \cite{voit} one expects that hole-hole interactions
drive the system toward a CDW phase, which would become insulating in
the presence of any amount of disorder \cite{nilsPRL}. 
Moreover, interactions
cause the density of states vanish at the chemical potential following 
a non universal power of the interaction parameter \cite{voit}. 

\subsection{Stripes, magnetism and kinetic energy release}

At half-filling, cuprates are antiferromagnetic Mott insulators
with the Cu atom carrying a spin of $1/2$. La$_2$CuO$_4$ is one of
the most striking examples of a layered antiferromagnetic 
Mott insulator. The N\'eel temperature, instead
of taking a value on the order of the planar magnetic exchange $J$ 
($\approx 1500$K), is approximately $300$K. This occurs
because of the low dimensionality of the system and indeed, 
an $O(3)$ invariant 2D Heisenberg model may order only 
at zero temperature due to the breaking of continuous symmetry 
(Mermin-Wagner theorem)
\cite{mermim-wagner}. However, the small inter-planar coupling, $J_{\perp}$
($\approx 10^{-4} J$), stabilizes antiferromagnetic order at finite 
temperature. Magnetism in these systems
is evident essentially across the entire phase diagram, although
long-range order is lost with only $2\%$ doping by Sr. 
Since the Sr atoms are located out of the CuO$_2$ planes and their effect is
to introduce holes, the doped holes are essential
for the destruction of long-range order. More importantly, it
is the minimization of {\it kinetic energy} of the holes which is the dominant mechanism
for the suppression of magnetism. In the following, we discuss several 
different ways to understand the importance of hole motion in these systems. 
   
A first indication for the importance of hole kinetic energy is given by
magnetic measurements for $x<0.02$ 
which observe the {\it recovery} of the magnetization
when the system is cooled below the so-called freezing temperature, $T_F(x)$ 
\cite{muSR}.
The staggered magnetization $M_S(x,T)$ vanishes at the N\'eel temperature 
$T=T_N(x)$ and is a smooth function for $T_F(x)<T<T_N(x)$. However for $T<T_F(x)$ the 
magnetization seems to recover to the full value expected at 
$x=0$ and $T=0$. This effect can be ascribed to localization of the
holes after which they affect the magnetization only 
locally. Exactly where the localization of holes occurs remains unresolved.
However, soft X-ray (oxygen K-edge) absorption experiments indicate that
the holes are probably in the oxygen sites \cite{xsoft}. NMR
experiments appear to indicate that holes would localize preferentially 
close to the charged Sr atoms for electrostatic reasons \cite{hammel}. However,
because the system is annealed as temperature is reduced, it is 
quite possible that the unscreened Coulomb interaction between holes
plays an important role in the localization process. If this is the
case, localized stripe patterns may form even at low doping, although
disorder effects from Sr doping are very strong in this region of the phase diagram
and one would expect any stripe pattern to be random in the CuO$_2$ planes 
\cite{branko}.
Recent neutron-scattering experiments at low doping find diagonal 
incommensurate peaks in the magnetic sector \cite{SG}. However, as no
charge peak has yet been observed in the spin-glass regime, these measurements
may be interpreted within the stripe model but the question remains open: 
the antiferromagnetic peaks could also be interpreted as the formation
of a spiral phase \cite{nils_glass}. In order to resolve this question unequivocally 
neutron-scattering experiments in samples heavily doped with spin-zero impurities 
such as Zn are required. If, as expected for a spiral state \cite{nils_glass}, 
the slope of the incommensurability as a function
of Sr concentration $x$ changes by a factor $(1 - 2 z)$, where $z$ denotes
the Zn concentration, the stripe hypothesis would be excluded in the spin-glass
regime.

Independent of the pattern of localization, holes may be
localized either on the O or the Cu sites. 
If hole localization occurs at the O sites, one would expect a spin-glass phase
to be observed at temperatures below $T_F(x)$ because a localized hole
at an O site liberates one spin (configuration $p^5$) which frustrates
the antiferromagnetic coupling between neighboring Cu atoms \cite{Aharony}. 
If, on the other hand, holes are localized on the Cu sites, the 
magnetization of the system would be reduced by one quantum of spin. Presumably,
because of the delocalization of the hole wave function between O and Cu
atoms the two effects can occur simultaneously \cite{zhang-rice}. The key feature
of these experiments is that they indicate the importance of the hole
motion for the destruction of long-range order. 

The rapid suppression
of magnetism with hole doping can be contrasted with the slow suppression of
antiferromagnetic order when Cu is replaced by a non-magnetic atoms such
as Zn or Mg \cite{vajk}. In this case, long-range order seems to be lost 
only at $41\%$ doping, that is, at exactly the classical percolation 
threshold for a 2D Heisenberg system. As has been shown in recent theoretical 
and numerical studies of the diluted {\it quantum} Heisenberg model, magnetic 
order seems to disappear only close to the
classical percolation threshold even in the quantum system \cite{sasha_afm}. 
These results have been investigated experimentally and in fact the
quantum fluctuations introduced by the dopants apparently are not sufficient 
to produce a quantum critical point (QCP) below the value of classical
percolation \cite{vajk}.  Although the comparison
between the problem of hole doping and Zn doping 
is not at first obvious, from the
chemical point of view Zn introduces a static hole in the Cu
plane because it has the same valence, but also has one extra proton. 
The situation on doping by Zn is therefore similar to the problem of holes
localized at Cu sites. Indeed the recovered magnetization in the 
hole-doped case, $M(x,0)$, is very close to the value which one 
would obtain on replacing Cu by a density $x$ of Zn atoms. 
Thus, further confirmation is obtained that the hole kinetic 
energy is the driving force behind the suppression of antiferromagnetic
order in these systems.

\section{Phenomenological models: stripes and antiferromagnetism}

In the previous sections we have argued that the kinetic energy of holes
is fundamental for understanding the stripe phenomenon 
in cuprates. However, dimensional crossovers are very difficult to measure
experimentally. Phase transitions, on the other hand
are easy to observe, because they produce strong effects in the thermodynamic
properties. Because Mott insulators are usually antiferromagnetic one may 
ask if such dimensional crossovers in the hole motion affect the
antiferromagnetic phase? In the previous section we provided evidence that the kinetic
energy of the holes is responsible for the destruction of antiferromagnetism
in these systems when the hole concentration is of order of $0.01$-$0.02$.
How is it possible that such small doping levels can destroy a robust
antiferromagnetic phase, with a N\'eel temperature which is of order $300 $K?

A possible way to understand the effect of the hole motion is to consider the
formation of infinitely long stripes. 
The first obvious effect is a breaking of the spatial rotational symmetry.
In the antiferromagnetic phase the spin rotational symmetry is also broken, 
indicating that both symmetries must be broken in the ground state of 
a striped antiferromagnetic phase. As a consequence the Goldstone modes
associated with the broken symmetries must carry information about 
them. For an ordinary (non-striped) antiferromagnetic phase these 
are spin-wave modes characterized by an energy
dispersion $E({\bf k}) = c_s k$, where $c_s$ is the spin-wave velocity,
and a spin-stiffness $\rho_s$ associated with the twist of the order
parameter \cite{chn}. In a striped antiferromagnetic phase the Goldstone modes 
remain spin waves, but because of the broken rotational symmetry their  
dispersion is different if the mode propagates
along the direction of broken symmetry or perpendicular to it, {\it i.e.,}
the energy dispersion is not circularly symmetric and $E(k_{||},k_{\perp})
= \sqrt{c_{||}^2 k_{||}^2 + c_{\perp}^2 k_{\perp}^2}$, where $||$ and $\perp$
refer respectively to the directions parallel and perpendicular to the 
stripes. 
At wave lengths longer than the stripe separation
$\ell$ and energies lower than the first spin-wave gap due to the folding of 
the Brillouin zone, this kind of dispersion is guaranteed by the nature of the 
broken symmetries.

In an ordinary antiferromagnet the spin-wave velocity is simply related
to the lattice spacing $a$ and the exchange constant $J$ by $c_s = S J a^d$.
Thus, a striped antiferromagnet may be modeled simply 
by assuming that the only effect of the stripes is to
introduce anisotropy in the exchange constants. Let us consider the case
of a spatially anisotropic Heisenberg model with exchange constants
$J_x$ and $J_y$ in the $x$- and $y$-directions, respectively, in which case
the spin-wave velocities in each direction are given by \cite{hone} 
\begin{eqnarray}\nonumber
c_y^2 = 2 S^2 a^2 J_y (J_x+J_y)
\\
c_x^2 = 2 S^2 a^2 J_x (J_x+J_y).
\label{velocities}
\end{eqnarray}
Microscopically, one may regard 
the stripes as causing local modification of the exchange across
an antiphase domain wall from $J$ to a value $J'$($<J$). This 
alteration of $J$ leads to a macroscopic change in the values of
the exchange constants in the same way that the introduction of impurities in a solid 
leads to an average change in the unit-cell volume. 

To relate $J$ and $J'$ to $J_x$ and $J_y$ is
not a trivial task. One possibility would be to solve the linear
spin-wave theory for the striped antiferromagnet and calculate the 
derivative of the spin-wave energy at the ordering vector ${\bf Q}$.
For stripes with a separation of $N_s$ lattice sites, this procedure requires the
solution of $N_s$ coupled differential equations. Besides being
computationally intensive, the solution would not provide significant insight into the
origin of $J'$ and would just exchange one phenomenological parameter
by another.

The simplest theory describing a striped antiferromagnet is the 
spatially anisotropic non-linear $\sigma$ model
\begin{eqnarray}
S_{eff} &=& \frac{1}{2}
\int_0^{\beta \hbar} d \tau \int dx \int dy
\left\{ S^2 \left[J_y \left(\partial_y \hat{n}\right)^2
+ J_x \left(\partial_x \hat{n}\right)^2 \right] \right.
\nonumber
\\
&+& \left.
\frac{\hbar^2}{2 a^2 (J_x+J_y)} \left(\partial_{\tau} \hat{n}\right)^2
\right\},
\label{effa1}
\end{eqnarray}
where $\hat n$ is a unit vector field. The symbols have been chosen to
suggest the continuum limit of an underlying effective {\it integer-spin}
Heisenberg Hamiltonian on a square lattice \cite{haldane_gap}. 
The underlying anisotropy parameter is the
ratio of the two exchange constants, or of the two velocities,
\begin{equation}
\alpha = J_x / J_y.
\label{anisotropy}
\end{equation}
The value of $\alpha$ characterizes the theory, but its exact dependence on
microscopic parameters is not easy to derive. 

We proceed by making use of well-established techniques to
analyze the behavior of the field theory described by the action
(\ref{effa1}) to predict the physical properties of
the system of interest.  It is useful to reexpress (\ref{effa1})
symmetrically by a dimensionless rescaling of variables
$x'= (\alpha)^{-1/4} x \Lambda$,
$y'= (\alpha)^{1/4} y \Lambda$ ($\Lambda\sim 1/a$ is a momentum cut-off), 
$\tau'= \sqrt{2 (J_x+J_y) \sqrt{J_x J_y}} S a  \tau/\hbar$.
The effective action (\ref{effa1}) becomes
\begin{equation}
S_{eff} = {\hbar\over (2 g_0)} \int_0^{\hbar \Lambda \beta c_0}
d \tau' \int dx' \int dy'
\left(\partial_{\mu} \hat{n}\right)^2,
\end{equation}
where $\mu$ denotes $ x',y'$, and $\tau'$,
\begin{equation}
g_0(\alpha) = \hbar c_0 \Lambda/\rho^0_s =
\left[2 (1+\alpha)/\sqrt{\alpha}\right]^{1/2}(a \Lambda)/S
\label{go}
\end{equation}
is~the~bare~coupling~constant, $c_0= [2 (J_x+J_y) \sqrt{J_x J_y}]^{1/2}
(a S)/\hbar$
the spin wave velocity and
\begin{eqnarray}
\rho^0_s = \sqrt{J_x J_y} S^2
\label{rhos}
\end{eqnarray} 
the classical spin stiffness of the rescaled
model. The original anisotropy is now contained in the limits of integration.

Notice that while $\alpha$ depends on the ratio $J_x/J_y$, the
spin stiffness depends on the product $J_x J_y$. Thus, given a
microscopic model where $J_x$ and $J_y$ are expressed in terms
of microscopic quantities the field theory is well defined. 
Unfortunately no calculations yet exist for the
microscopic form of these quantities, and certain 
assumptions are required concerning their behavior. If the spatial rotational symmetry
is broken at the macroscopic level, that is, one has infinitely
long stripes in the $y$ direction, a simple choice would 
be $J_y = J$ and $J_x = \alpha J$, whence
\begin{eqnarray}
\rho^0_s = \sqrt{\alpha} \rho_I
\label{zaanen}
\end{eqnarray}
where 
\begin{eqnarray}
\rho_I = J S^2
\end{eqnarray}
is the spin stiffness of the isotropic system. This choice is valid only
when the system is composed of a mono-domain of stripes \cite{vanduin}. 
If the system is broken into domains, in which case the rotational symmetry
is broken micro- but not macroscopically the choice (\ref{zaanen}) may
not be the most appropriate. At sufficiently long wave lengths the
system is essentially isotropic and therefore the spin stiffness
is the same in all directions, \cite{hone}
\begin{eqnarray}
\rho^0_s = \rho_I,
\label{hone}
\end{eqnarray}
which is obtained by choosing $J_x = \sqrt{\alpha} J$ and 
$J_y = J/\sqrt{\alpha}$. We note that these choices
are essentially arbitrary and are based on qualitative expectations concerning
the nature of the correlations at very long wave lengths. 
One may show that the choice (\ref{hone})
is appropriate very close to the antiferromagnetic phase ($x < 0.02$)
where the breaking of the system into domains is quite probably because of 
disorder effects \cite{hone}. However, at larger doping ($x =1/8$)
the choice (\ref{zaanen}) is more appropriate because long-range
stripe order is observed \cite{vanduin}. With these two parameterizations 
one may analyse the problem and calculate
physical quantities for comparison with experiments.
One general consequence of the anisotropy introduced by the presence of
stripes is a growth of quantum fluctuations because of
the reduction of effective dimensionality. In fact,
by using large $N$ methods and RG calculations, one may demonstrate that
the effective spin stiffness is given by
\begin{equation}
\rho_s(\alpha) = \rho_s^0(\alpha)\left[
1 - \frac{g_0(1)}{g_c(\alpha)}
\right],
\label{stiff}
\end{equation}
where
\begin{eqnarray}
g_c(\alpha) &=&
 \sqrt{8} \pi^{2} \sqrt{\alpha/(1+\alpha)}  \left\{
 \ln\left(\sqrt{\alpha}+\sqrt{1+\alpha}\right)\right.
\nonumber
\\
&+& \left. \sqrt{\alpha}\ln[(1+\sqrt{1+\alpha})/\sqrt{\alpha}]
\right\}^{-1}
\label{gca}
\end{eqnarray}
is a critical coupling constant. 
Notice that (\ref{stiff}) is
reduced from its classical value $\rho_s^0(\alpha)$ for fixed anisotropy
$\alpha$, and that it vanishes at some critical value $\alpha_c$, whence 
$g_c(\alpha_c)=g_0$. Hence, as a function of the anisotropy, the model exhibits
quantum critical point where the system undergoes a quantum
phase transition from an ordered N\'eel state to a paramagnetic phase. 
The loss of antiferromagnetic order at $x=0.02$ can thus be considered 
as a consequence of the enhancement of quantum fluctuations 
due to the presence of stripes. 

As explained previously, these considerations are valid when the holes move
along the stripes and modify the exchange constant across the stripe.
However, when localization occurs at low temperatures (as observed in the
recovery of magnetization in NQR experiments) the stripes essentially
cease to exist and the system undergoes a phase transition into a 
spin-glass phase. The simplest way to understand this phase is to consider 
hole localization at the oxygen sites with consequent liberation of one spin 
$1/2$ \cite{Aharony}. This spin frustrates the antiferromagnetic order, because the
superexchange interaction of the O spin with the neighboring Cu spins
is antiferromagnetic. This problem may be treated by considering the 
O spin as a classical localized dipole moment \cite{nils_glass}. 

Finally, the theoretical results concerning the existence of stripes and
the appropriate model for describing them can be summarized as follows: 
the proposal that the doped $t$-$J$ model undergoes a phase separation is supported
by variational arguments \cite{emery2}, diagonalization on small clusters 
\cite{emery2}, and Green-function quantum Monte Carlo calculations 
\cite{manou}.
On the other hand, several quantum Monte Carlo calculations \cite{QMC},
series expansions \cite{series}, exact diagonalization \cite{ED}, 
and DMRG calculations \cite{DMRG}
yield results contradicting these claims and supporting the stripe
picture. In order to gain more insight into the problem, we begin by considering
the simplest possible case, namely, the single-hole problem. 

\section{Phenomenological models: Transverse fluctuations and pinning of stripes}

\begin{figure}[ht]
\vskip-.6in
  \centerline{\epsfxsize=0.70\columnwidth\epsffile{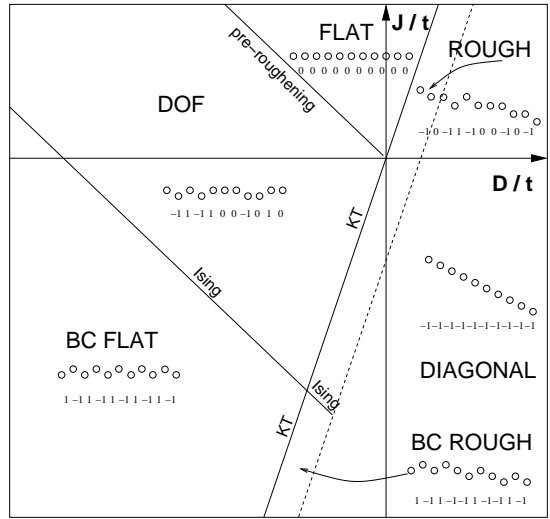}}
\vskip-.6in
\caption{Sketch of the phase diagram for the spin-1 chain. The stripe 
configurations represented by circles, and the corresponding values 
of $S^z$ are shown below. There are six different phases: 
1) a gapped flat phase, corresponding to straight stripes ($S_z = 0$); 
2) a gapless rough phase (spins equal to zero and $\pm1$  
distributed randomly); 3) a gapped bond-centered (BC) flat phase, which has a 
long-ranged zig-zag pattern (periodic alternation of $S_z = 1$ and 
$S_z = -1$); 4) a gapless BC rough phase with a zig-zag pattern 
(antiferromagnetic correlations with disordered 
$S_z=\pm 1$ but no $S_z=0$ states); 
5) a diagonal stripe phase, corresponding to a ferromagnetic 
state in the spin language; 6) a gapped disordered flat phase (DOF),
where a kink $(S_z = +1)$ is followed on average by an anti-kink $(S_z = -1)$,
but with some $S_z = 0$ states in between.}
\label{DiagramS1}
\end{figure}

While much theoretical effort has been concentrated on determining
whether stripes are the ground state of models such as the Hubbard
and the $t$-$J$ models, which are supposed to describe high-$T_c$ 
superconductors, a parallel research direction has also developed which
consists of phenomenological studies of the striped phase. 
In this case, one assumes the existence of stripes
and discusses further aspects such as their effect on the antiferromagnetic
state (see previous section). Motivated by issues such as the static or
fluctuating nature of stripes and the mechanism of stripe pinning, 
a phenomenological theory for the pinning of stripes has been 
developed. Zaanen {\it et al.} have related transverse
stripe fluctuations to the restricted solid on solid model (RSOS) which describes 
the growth of surfaces \cite{eskes}. In a simplified form of the model, the transverse 
kink excitations of stripes are mapped to a quantum spin-1  
chain model \cite{eskes}, whose Hamiltonian is 
\begin{equation}
H = \sum_n \left[ - t \left( S_n^x S_{n+1}^x +S_n^y S_{n+1}^y
\right) - D S_n^z S_{n+1}^z + J (S_n^z)^2 \right].
\label{S1}
\end{equation}
Here, $t$ has the role of a hopping parameter
for transverse kinks, $J$ controls the density of kinks
and the $D$ term represents a nearest-neighbor interaction of kinks.
The spins take the values $S_n^z = 0, \pm1$, where $+1$ and $-1$ are
associated respectively with stripe kinks and anti-kinks and
$S_n^z = 0$ describes unperturbed (flat) segments. 
The full phase diagram for this problem was determined
numerically by den Nijs and Rommelse \cite{denN} after earlier
calculations by Schulz \cite{Schu}, who treated the
spin-1 problem as two coupled spin-1/2 chains. We have recently reanalysed the
calculations of Schulz \cite{Nils3} and derived the correct phase diagram 
from this formalism, which agrees with the one obtained in Ref.\ \cite{denN}
(see Fig.\ \ref{DiagramS1}). Six different phases can be identified,
depending on the values of the $D$ and $J$ parameters. If $J$ is positive, 
the last term of Eq.\ (\ref{S1}) determines that $S_n^z = 0$ and the stripe is
straight (flat phase). If $J$ is negative, both values 
$S_n^z = \pm 1$
are equally favorable with respect to the $J$ term and the $D$ term 
determines the value of $S_n^z$. If $D$ is positive, nearest-neighbor segments 
prefer to be similar and therefore $S_n^z$ and $S_n^{z + 1}$
will have the same sign. This gives rise to the ferromagnetic phase (diagonal
stripes), with a sequence of kinks or anti-kinks. On the other
hand, if $D$ is negative, $S_n^z$ and $S_n^{z + 1}$ prefer to have opposite
signs, and the stripe will be bond-centered and flat, with a ``zig-zag'' shape 
(a kink follows an anti-kink and {\it vice versa}). In addition, both the flat 
and the bond-centered flat phases, which are gapped, can undergo a 
Kosterlitz-Thouless transition to 
gapless rough or bond-centered rough phases, respectively. 
The sixth phase, which was not identified by Schulz, corresponds to 
a gapped, disordered, flat phase (DOF), (Fig.\ \ref{DiagramS1}).
In contrast to the flat phase, this phase has a finite density
of kinks and anti-kinks, which are positionally disordered,
but have an antiferromagnetic order in the sense that a kink $S_z=1$ is on 
average followed by an anti-kink $S_z=-1$ (rather than another kink),
with any number of $S_z=0$ states in between them. The DOF phase is
the valence-bond phase which is responsible for the Haldane gap.

In the limit of negligible nearest-neighbor interaction ($D \sim 0$), 
the above model can be related to the $t$-$J$ model \cite{cris} by considering 
the transverse dynamics of a single vertical stripe in a frozen N\'eel 
background. The transverse dynamical properties of holes are 
described by the $t$-$J_z$ model, and they move under the
condition that the horizontal separation between two
neighboring holes cannot be larger than one lattice constant. 
Notice that this condition does not restrict the motion of the stripe:
it may still perform excursions very far from the initial
straight-line configuration, but the line cannot be broken.
In this case the $t$-$J$ Hamiltonian describing hole
dynamics may be mapped onto a spin-1 chain Hamiltonian
analogous to Eq.\ (\ref{S1}), but with $D = 0$ \cite{cris}. 
A duality transformation of the initial quantum string Hamiltonian maps 
the problem onto one describing a 1D array of Josephson
junctions \cite{dimashko}, which is known to exhibit an 
insulator-superconductor transition at $(t/J)_c =2 / \pi^2$ \cite{BD}. 
This Kosterlitz-Thouless transition represents the
unbinding of vortex-antivortex pairs in the equivalent $XY$ model, which
translates to a roughening transition for the stripe problem. In this way, 
the transition between the flat and rough phases along the vertical line
$(D = 0)$ of the phase diagram could be determined precisely by analytical means 
\cite{dimashko}.   

These models are related to the sine-Gordon model, and by
studying the spectrum of the quantum
string in a Hilbert-space sector of zero topological charge, 
the meaning of the transition in the ``string'' language may be clarified. 
At $(t/J)_c$ the (insulating) pinned phase,
which has an energy spectrum with a finite gap,
turns into a (metallic) depinned phase where the spectrum becomes gapless
\cite{dimashko}.
This procedure allows the connection of two important and different classes of
problems, namely the transverse dynamics of stripes in doped antiferromagnets
and a system with the well-known properties of the sine-Gordon model.

In all the models discussed hitherto, the pinning potential arises from the
discrete nature of the lattice. However, the introduction of holes into the 
MO$_2$ planes (M = Cu or Ni) is not the only consequence of  
doping a Mott insulator such as La$_2$MO$_4$.
Doping an antiferromagnetic insulator also introduces disorder into this
material due to the presence of counterions, which act 
as attractive centers for holes. Doping with divalent atoms, such as 
Sr$^{2+}$ produces quenched disorder, because the ionized dopants are 
located randomly between the CuO$_2$ planes. In contrast, 
doping with excess oxygen generates annealed disorder. 
Indeed, oxygen atoms have a low
activation energy and remain mobile down to temperatures of 200-300 K.

In order to account for the random pinning potential provided by the Sr
atoms in nickelates and cuprates, one may add a disorder potential
to the previous phenomenological Hamiltonian (\ref{S1}) with $D = 0$. 
This allows a determination of 
the influence of both disorder and lattice effects on
the striped phase of cuprates and nickelates. We consider the
problem of a single stripe along the vertical direction confined in a box of size 
$2 \ell$, where $\ell$ denotes the stripe spacing. 
The system is described by the 
phenomenological Hamiltonian
\begin{equation}
\hat{H}=\sum_n \left[ - 2 t \cos \left( \frac{\hat{p}_n }{\hbar}\right) +
J  \left(\hat{u}_{n+1}-\hat{u}_n \right)^2 +
V_n (\hat{u}_n)\right],
\label{Ham1}
\end{equation}
with $t$ the hopping parameter, $\hat{u}_n$ the displacement of
the n-th hole from the equilibrium (vertical) configuration, $\hat{p}_n$
its conjugate transversal momentum, $J$ the stripe stiffness, and
$ V_n (\hat{u}_n)$ an uncorrelated disorder potential satisfying 
$\langle V_n (u) V_{n'} (u') \rangle_d = d \delta (u - u') \delta_{n,n'}$, where
$\langle ... \rangle_d$ denotes the Gaussian average over the disorder ensemble and
$d$ is the inverse of the impurity scattering time.
Eq.\ (\ref{Ham1}) may be straightforwardly related to Eq.\ (\ref{S1}) by noting
that $S_n^z = u_{n + 1} - u_n$ and that the hopping terms $S_n^x$ and $S_n^y$ are
connected to translation operators, which can be written in the momentum 
representation $p_n$ as $\tau_{\pm} = {\rm e}^{\pm i p_n / \hbar}$ 
\cite{dimashko}. We are considering the lattice parameter $ a = 1$. 

A dimensional estimate provides the dominant features of the phase
diagram.
At large values of the hopping constant $t \gg J$, the first term may be expanded
as $-2t\cos(\hat{p}_n /\hbar) \sim const. + t (\hat{p}_n /\hbar)^2$.
In the case of no impurity potential, $ V_n (\hat{u}_n)= 0$, hole  
dynamics is governed by the competition between the kinetic term
$t (k_n )^2$, which favors freely mobile holes, and the elastic one, $J 
(\hat{u}_{n+1}-\hat{u}_n)^2$, which acts to keep them together.
When the confinement is determined by the lattice pinning potential, 
the average hole displacement $\hat{u}_{n+1}-\hat{u}_n $ is of order $1$ 
(the lattice constant is unity) so 
the wave vector $k_n \sim 1$. A transition from the flat phase, with the 
stripe pinned by the underlying lattice, to a free phase is then expected 
at $t/J \sim 1$.

We now consider the opposite limit of strong pinning by impurities.
In this case, the potential provided by the lattice is irrelevant and
the typical hole displacement is on the order of the separation between
stripes, $1/k_n \sim \hat{u}_{n+1}-\hat{u}_n \sim \ell$. By
comparing the kinetic
$t (1/\ell)^2$ and the elastic $J (\ell)^2$ terms, we observe that a
transition should occur at $t/J \sim (\ell)^4$.

\begin{figure}[ht]
  \centerline{\epsfxsize=0.70\columnwidth\epsffile{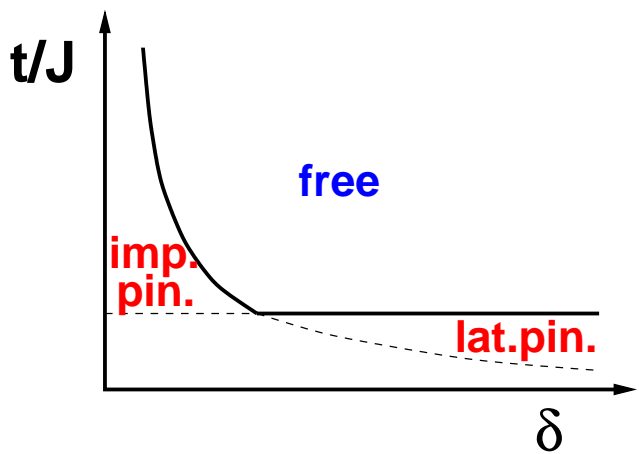}}
\caption{
Zero-temperature pinning phase diagram of the stripe phase in the
presence of lattice and impurity pinning.  Three phases can be identified: 
a quantum membrane 
phase with freely fluctuating Gaussian stripes, a flat phase with the stripes 
pinned by the lattice, and a disorder-pinned phase \cite{nilsPRL}.}
\label{pinning}
\end{figure}

Indeed, by deriving the differential renormalization group (RG) equations
to lowest nonvanishing order in the lattice and disorder parameters, one
obtains a set of flow equations \cite{nilsPRL}, which indicate that the transition
from the flat (lattice-pinned) to the free phase occurs at 
$(t/J)_c = 4/\pi^2$, and
the transition from the disorder-pinned to the free phase occurs at 
$(t/J)_c = (36/\pi^2) \ell^4$. The pinning phase diagram of the striped phase 
is shown in Fig.\ \ref{pinning}, in which $\delta = 1/ 2 \ell$. 

By comparing these results with recent measurements on nickelates
and cuprates one concludes that nickelates occupy the lower left corner of
the phase diagram, i.e., they have static stripes which are pinned by the lattice
and by the impurities. By contrast, cuprates are characterized by freely fluctuating
stripes and so appear in the upper right corner. 
An appropriate treatment of the striped phase in cuprates must therefore include
stripe-stripe interactions and the model becomes similar to that for 
a 2D fluctuating membrane \cite{nilsPRL,Scheidl}.
In this phenomenological framework, both nickelate and cuprate 
materials can be 
understood in a unified way, the difference between them being simply the 
parameter $t/J$ which measures the strength of quantum fluctuations.

Although the number of holes is intrinsically connected with the number
of pinning centers, recent experimental developments show that it is
possible to control these two parameters independently. Co-doping
the superconducting cuprate material LSCO with Nd or Zn 
increases the disorder without modifying the number of charge carriers 
\cite{tran,koike,xiao}. On the other hand, 
growing the superconducting film on a ferroelectric substrate and using an 
electrostatic field as the control parameter allows the number of charge 
carriers in the plane to be altered for a fixed Sr concentration $x$ \cite{ahn}. 
This class of experiments constitutes an important step towards the control of a
normal metal-superconducting transition.

One may then investigate the stripe pinning produced 
by Zn and Nd co-doping \cite{AIP}. 
The two dopants play fundamentally different roles in the
pinning process. Nd, as with other rare-earth co-doping, induces a
structural transition which produces a correlated
pinning potential trapping the stripes in a flat phase. The situation is
analogous to the pinning of vortices by columnar defects or screw dislocations 
\cite{vortices}.
In this case transverse fluctuations are strongly suppressed, long-range order 
is achieved and thus the incommensurate peaks observed by 
neutron diffraction become sharper after the introduction
of the co-dopant, as observed experimentally \cite{tran}. On the other hand, in-plane
Zn- and Ni-doping provide randomly distributed point-like pinning centers,
similar to oxygen vacancies in the vortex-creep problem.
Within the model in which a stripe is regarded as a quantum elastic
string, the effect of randomness is to ``disorder'' the string, promoting
line-meandering, destroying the 1D behavior and broadening
the incommensurate peaks. A perturbative treatment of the RG equations
discussed previously show at the next higher order that this kind of pinning 
is relevant only in under-doped systems, in agreement with experiments \cite{hiro}. 

Finally, it is essential to go beyond the studies of transverse stripe excitations
to consider the coupling between longitudinal and transverse modes. 
Longitudinal modes may be described by a Luttinger-liquid Hamiltonian, and
the coupling between longitudinal and transverse fluctuations was investigated 
using bosonization \cite{Nils3}. One finds that a longitudinal CDW instability 
can arise if the stripe is quarter-filled and the underlying lattice potential 
has a zig-zag symmetry. This result has shed additional light on the 
connection 
between the formation of a LTT phase and the subsequent appearance of
charge order in high-$T_c$ cuprates (Fig.\ \ref{Ichi}).

Experimentally, the suppression of superconductivity in LSCO co-doped
with Nd (Fig.\ \ref{Ichi}), and also the upturn of the
resistivity in the normal phase, are correlated with a structural
transition from the LTO to the LTT phase. Indeed, at $x = 1/8$ the
charge ordering temperature $T_{co}$ reaches its highest value and the 
superconducting temperature $T_c$ shows a local minimum. Neutron diffraction  
experiments indicate that in the underdoped phase of LSCO the stripes
are quarter-filled \cite{tran}. The formation of the LTT phase favors
a zig-zag symmetry of the transverse stripe degrees of freedom.
Thus, below $T_{LTT}$ the CDW instability discussed
above becomes relevant and stabilizes a bond centered string with
zig-zag symmetry. If the stripe spacing is exactly commensurate, as
at $x=1/8$, a long-range-ordered CDW (Wigner crystal)
can form, leading to the suppression of $T_c$. On the other hand,
for incommensurate doping values, solitonic modes are
present in the stripe which prevent long-range charge order. 
Longitudinal charge order has hintherto not been observed. Nonetheless,
the upturn of the in-plane resistivity below $T_{LTT}$
suggests proximity to an insulating phase inside the LTT phase
of underdoped cuprates, which is likely to be the bond-centered zig-zag 
stripe \cite{Nils3}.

\section{Experimental detection of stripes in YBCO and BSCCO}

An important question at this point is whether charge stripes are
peculiar to the lanthanates or a generic feature of all the
cuprates. The answer is not yet completely settled, and further
experiments are required to achieve an unambiguous conclusion. 
Measurements on YBCO and BSCCO compounds begin to provide a 
comprehensive picture.

Inelastic neutron scattering experiments recently performed by Mook 
{\it et al.} on YBCO$_{6.6}$ \cite{mook1}, which corresponds to a hole 
doping of $x = 0.10$,  
detected a dynamical incommensurability in the magnetic sector of 
$\delta = 0.105 \pm 0.01$,
which is exactly the value measured by Yamada {\it et al.} for the
corresponding charge concentration in LSCO \cite{yama}. Later, magnetic 
incommensuration was also observed at $\delta = 0.0625$ in YBCO$_{6.35}$, 
which corresponds to a doping $x = 1/16$ \cite{mook2}. In addition, measurements
of phonon broadening at a wave vector consistent with the stripe picture 
(twice the magnetic wave vector) confirmed
the previous results for YBCO$_{6.6}$ and YBCO$_{6.35}$ \cite{mook2}. 
Eventually, a static charge order peak has been observed in a 21g crystal of 
YBCO$_{6.35}$, at a wave vector which is exactly double the 
dynamical magnetic incommensuration, 
$2 \delta = 0.127$, as expected within the stripe picture \cite{mook4}. Although 
the charge peaks are small, 6 orders of magnitude below 
the strongest crystal Bragg peak, their existence is undeniable. However, it
should be emphasized that no charge order has been observed in YBCO$_{6.5}$ and 
YBCO$_{6.6}$, so the situation is not yet resolved.

Another important feature measured recently in YBCO is the 1D nature of the 
stripes \cite{mook3}. In a 4g crystal of detwinned YBCO$_{6.6}$, one could
observe not four, but only two incommensurate magnetic peaks (the second set
of perpendicular peaks have nearly vanished, because the sample was almost
completely detwinned). The results suggest
stripes aligned along the $b-$axis, in agreement with far-infrared 
spectroscopic measurements by Basov {\it et al.} \cite{basov}, which indicate that
the superfluid density is larger along the $b-$direction. This behavior cannot
be attributed to the chains, because in underdoped materials the missing atoms
in the chains would inhibit the chain contribution to superconductivity.

As a conclusion, one could state that neutron scattering experiments in YBCO
at several different doping concentration \cite{mook5} confirm the stripe picture
sketched for LSCO and reveal the universality of the previous results.

Concerning BSCCO, the majority of experimental results are obtained 
from scanning tunneling microscopy (STM) and spectroscopy (STS) 
[7-12]. 
The advantage of STM lies in its ability to measure simultaneously, with atomic
resolution, both the surface topography and the local density of states (LDOS)
of a material. The topographic image can be realized due to the exponential
dependence of the tunneling current $I$ on the separation of tip and
sample. In addition, the differential conductance $G = dI/dV$, where $V$ is
the sample bias voltage, is proportional to the LDOS of the sample at the tip
location. 

Low-$T$ STS in BSCCO samples ($T_c = 87$K) revealed the existence of a large 
number of randomly distributed regions, with characteristic lengthscales
of order 30 \AA, which have anomalous LDOS
features. These features were initially referred to as quasiparticle scattering
resonances (QPSRs) and were thought to be due to quasiparticle scattering from
atomic-scale defects or impurities \cite{Hudson}, but later measurements performed
in Zn-doped samples indicated that these inhomogeneities do not originate from
impurities. The locations of Zn impurities were identified from the zero-bias 
conductance map and the LDOS map taken simultaneously at the same 
location showed no correlation between the intensity of the integrated
LDOS and the location of the Zn impurities \cite{Pan}.

Spatial variations of the tunneling spectrum and of the superconducting gap
can be observed in pure as well as in impurity-doped BSCCO samples, and seem to be
intrinsic to the electronic structure. The gap ranges from 25meV to 65meV
and has a gaussian distribution \cite{Pan}. The average gap is very similar to that 
reported previously from tunneling measurements. The spectra obtained at points
with larger integrated LDOS exhibit higher differential conductance, smaller
gap values and sharper coherence peaks, which are the characteristic features
of spectra taken in samples with high oxygen doping concentration. This 
observation suggests the interpretation
that the inhomogeneities may arise from differences in local oxygen concentrations 
\cite{Pan}. 

Recent experiments in underdoped BSCCO indicate that these high-$T_c$ materials are
granular superconductors, with microscopic superconducting grains separated
by non-superconducting regions. By doping the material with Ni impurities,
it was observed that the position of Ni atoms coincide with regions of small gap
($\Delta < 50$ meV). In underdoped BSCO these small-gap regions, which have  
large $G(\Delta)$ are separated by percolative regions with large gap $\Delta$ and 
low $G(\Delta)$ \cite{Lang}. 

Qualitatively new information is provided by STS measurements in magnetic field
up to $7.5$T. The quasiparticle states generated by vortices in overdoped BSCCO show
a ``checkerboard'' pattern with a periodicity of four unity cells \cite{hoff}, in 
agreement with the charge periodicity expected within the stripe picture \cite{tran}.
Indeed, the magnetic spatial periodicity previously detected by neutron
scattering in overdoped
LSCO in the presence \cite{Lake} or absence \cite{yama} of a magnetic field is
eight unit cells, exactly twice the charge spacing. Shortly after
these measurements, the $4a$ periodicity was observed by Kapitulnik {\it et al.} 
in nearly optimally doped BSCCO {\it without} magnetic field \cite{Kapitulnik}.
Transformation of the real-space data to reciprocal space showed a periodicity 
of $4a$ in the randomly distributed regions with anomalous LDOS,
which was manifest in four distinct peaks in reciprocal space 
\cite{Kapitulnik}. Two peaks corresponding to a periodicity of $8 a$ along the 
diagonal due to the periodically missing line of Bi atoms in the BiO
plane were also seen clearly, confirming the sample quality.
However, later studies by Davis {\it et al.} have cast doubt on the
stripe interpretation: by analyzing the energy dependence of the wave vectors 
associated with the modulation, they have argued that the checkerboard LDOS
modulation in BSCCO is an effect of quasiparticle interference, and not
a signature of stripes \cite{Davis}. Thus there is currently no agreement 
concerning the interpretation of the STM data. 

In summary, one may state that the presence of stripes
is now firmly established in LSCO (static in Nd-doped LSCO, dynamical in pure
LSCO). Recent neutron scattering experiments indicate that they are also present 
in YBCO. However, the magnetic YBCO stripes seem to be dynamical, whereas the 
charge ones are static. 
Concerning BSCCO, only STM measurements, which are just sensitive to charge
and have the disadvantage of being susceptible to surface defects and  
pinning, were performed hintertho. They suggest that a static charge stripe
is present in this compound, although many further measurements are required 
for a definitive understanding.

\section{Conclusions}

Doped antiferromagnetic insulators have recently attracted a great
deal of attention because many of the materials in this class exhibit novel
and interesting behavior. The cuprates, for example, 
become metallic at low doping concentration, and even superconducting
at relatively high temperatures, whereas other systems, such as nickelates,
show metallic behavior only at very high doping and are never superconducting.
Manganites, on the other hand, can display the phenomenon of giant magneto resistance.
Despite the different electric and magnetic properties exhibited by these
compounds, a common perovskite structure connects them. Moreover,
spontaneous symmetry breaking and stripe formation seems also to be a shared
feature.

Stripe-like ground states were first predicted from Hartree-Fock calculations
of the Hubbard model [49-53]. Later studies of the $t-J$ model also confirmed
the initial results \cite{DMRG}. However, this theoretical work attracted significant
attention only when Tranquada {\it et al.} detected the existence of static 
spin and charge order in Nd-doped LSCO 
\cite{tran}. The peaks measured by elastic neutron scattering were incommensurate with
pure antiferromagnetic order, and suggested that the system had undergone a 
phase-separation into 1D regions rich in holes (stripes), which were acting as 
domain walls in the staggered
magnetization. The fact that the magnetization changes phase by $\pi$ when crossing
a domain wall has the consequence that the magnetic periodicity is twice
the charge-stripe spacing. The associated incommensurability is therefore one half 
of the charge incommensurability, and the detection of both magnetic and charge 
incommensuration gave undeniable support to the stripe theory \cite{tran}.
Later inelastic neutron scattering experiments in pure LSCO showed that there is
a spin gap in these materials, but that incommensurate peaks can still be measured
at rather low energies. The actual value of the spin gap
depends on the doping. The incommensurability measured in these compounds is 
{\it exactly} the same as that obtained in Nd-doped LSCO, which is 
understandable considering replacement of La$^{3+}$ with Nd$^{3+}$ does not 
add charge carriers to the
system, but induces a structural transition that helps to pin the stripes.
The presence of stripes in LSCO has been confirmed by
several different experimental techniques. 
Inelastic
neutron-scattering experiments [17-23] indicate that incommensurate
peaks are also detectable in YBCO, with a linear dependence of the 
incommensurability as a function of doping concentration similar to that
observed in 
lanthanates. These results seem to indicate that the striped phase could be
a generic feature of cuprates, instead of a peculiarity of LSCO. The 
requirement of very large samples imposed by the neutron scattering renders 
experimental progress in the field very slow. Concerning BSCCO, at present 
only STM data is available and while these may support the existence of stripes 
\cite{hoff,Kapitulnik}, the question remains open. It is important to note here 
that STM is a surface probe, whereas neutron scattering 
measures bulk properties. Thus, surface impurities 
could play a dominant role in STM experiments, but a secondary one in neutron 
scattering. The years to come will hopefully show the truth behind all the
controversies.

From the theoretical side there has been steady progress in 
understanding stripe phases. Different approaches have been applied
to the problem with differing degrees of success. Mean-field theories
of different kinds \cite{italianos,vanhove,ash}, gauge theories \cite{zaa}, 
and quantum liquid crystal phenomenology \cite{kf} have been
employed to describe the stripe state. In this review we focused
primarily on studies based on the $t$-$J$ model, where DMRG calculations have
shown the presence of stripe phases \cite{sasha_stripe,uswhite}. These
studies have the advantage that rather few parameters determine the physics 
of stripe formation. 
Although this kind of approach describes very well the nature of the
stripes, it seems to indicate that the stripe state is essentially insulating and extra
degrees of freedom, such as phonons \cite{antonio}, may be required to explain
the experimental data in these systems \cite{egami,lanzara}.

Phenomenological models have also contributed to a significant evolution in 
understanding. Appropriate models for describing stripe
fluctuations have been developed, and analogies with other known systems
established \cite{eskes,cris,dimashko}. The effect of stripe pinning 
by impurities and by the underlying lattice, as well as the differing roles 
of rare-earth and planar impurities, has been clarified 
\cite{nilsPRL,Nils3,dimashko,Scheidl,AIP}.

An important task remaining for the coming years is to show, both experimentally  
and theoretically whether and how stripes are connected to superconductivity.
Systematic investigation of the different ways to suppress superconductivity
may yield answers to this complex question.

\begin{acknowledgments}
This work is the result of many hours of conversation and discussion with our 
colleagues and friends.
We are particularly indebted to A.~Balatsky, D.\ N.\ Basov, D.~Baeriswyl, 
A.\ Bianconi, 
A.\ Bishop, A.~Caldeira, E.~Carlson, A.~Chernyshev,  E.\ Dagotto, T.\ Egami,
E.~Fradkin, F.~Guinea, M.~Greven, N.~Hasselmann, D.~Hone, S.~Kivelson, A.\ Lanzara, 
G.\ B.\ Martins, R.\ McQueeney, B.\ Normand, S.\ H.\ Pan, L.~Pryadko,
J.\ Tranquada, S.~White, and J.~Zaanen. 
This review would not have been possible without their 
support and criticism. C.\ M.\ S.\ acknowledges finantial support from the 
Swiss National Foundation under grant 620-62868.00.
\end{acknowledgments}

\end{document}